\documentclass[amsmath,amsfonts,amssymb]{imsart}
\usepackage[dvips]{graphicx}

\begin{document}

\begin{frontmatter}
\title{Statistical Laws in Urban Mobility from microscopic GPS data in the area of Florence}
\runtitle{Statistical Laws in Urban Mobility}

\begin{aug}
\author{\fnms{Armando Bazzani} \ead[label=e1]{bazzani@bo.infn.it}},
\author{\fnms{Bruno Giorgini} \ead[label=e2]{}},
\author{\fnms{Sandro Rambaldi} \ead[label=e3]{}},\\
\author{\fnms{Riccardo Gallotti}\ead[label=e4]{}}
\and
\author{\fnms{Luca Giovannini} \ead[label=e5]{}}

\address{Physics of the City Laboratory, Physics Department and C.I.G.\\ University of Bologna, Italy,\\
INFN, sezione di Bologna, Italy \\ \printead{e1}}

\runauthor{A. Bazzani et al.}


\end{aug}



\begin{abstract}
The application of Statistical Physics to
social systems is mainly related to the search for macroscopic laws,
that can be derived from experimental data averaged in time or
space,assuming the system in a steady
state. One of the major goals would be to find a connection between the statistical laws to the
microscopic properties: for example to understand the
nature of the microscopic interactions or to point out the existence
of interaction networks. The probability theory suggests the
existence of few classes of stationary distributions in the
thermodynamics limit, so that the question is if a statistical
physics approach could be able to enroll the complex nature of
the social systems. We have analyzed a large GPS data base for single
vehicle mobility in the Florence urban area, obtaining
statistical laws for path lengths, for activity downtimes and for activity degrees. We show also
that simple generic assumptions on the microscopic behavior could
explain the existence of stationary macroscopic laws, with an universal function describing
the distribution. Our conclusion
is that understanding the system complexity requires dynamical data-base
for the microscopic evolution, that allow to solve both small
space and time scales in order to study the transients.
\end{abstract}


\end{frontmatter}

\section{Introduction}
Any statistical analysis of real systems is based on the Ergodic Principle for the microscopic
dynamics, that implies the relaxation towards steady states and the independence property
of elementary components. Even if the existence of microscopic
interactions is necessary for the system to evolve toward a statistical equilibrium, in this
state any particle moves independently from the others and all the particles are statistically equivalent (any particle
may be representative for the whole).
The thermodynamics laws that are derived from a statistical mechanics approach, concern some macroscopic
observables of the system, evolving adiabatically with respect the microscopic relaxation time (i.e.
we can consider the whole system in a almost equilibrium state), so that the effects of single particle
dynamics are conveniently described by means of stochastic processes.
As a consequence, there should exist a natural separation among macroscopic and microscopic space-time scales.
Indeed space and time scales are expected to be strictly correlated: to understand small scale phenomena
we need to solve short time scales and viceversa.
Nevertheless the statistical mechanics has a great success in describing evolution of macroscopic
systems and there is a strong effort to generalize the results for a non-equilibrium thermodynamics and
for application to complex systems\cite{balescu1975}. The statistical properties of social systems have
been recently considered under a different point of view due to the
possibility of recording large microscopic data sets\cite{brockmann2005,gonzalez2008}. The main
problem is what are the macroscopic effects of cognitive behavior for
"social particles". Indeed the cognitive behavior would imply the existence
of strong bidirectional interactions among the dynamics at different
space and time scales of the system\cite{gelder1998}. Emergence and
self-organization characterize the macroscopic states, but the
question is which macroscopic observables (if they exist) may enrol
the complex nature of the system. These variables may also play an
important role in the study of phase transitions and in the control parameters
definition.\par\noindent
In Italy GPS data on
individual vehicle paths are currently recorded for insurance reasons over a
sample $\simeq 2\%$ of the whole private vehicle
population\cite{octotelematcs,bazzani2007}. This data set gives the opportunity to study the
individual mobility demand in urban contexts. The GPS data set
contains the geographical coordinates, the time, the instantaneous
velocity and the path length of individual trajectories at positions
whose relative distance is of order $1\div 2$ km. Special signals
are recorded when the engine is switched on and off. We remark that
the data refer mainly to the private transportation mobility and
that, due to privacy legal problems, we do not have any knowledge on
the social features of individuals in the sample.\par\noindent
In this paper we analyze the statistical distributions of the path
lengths of individual trajectories, of the activity downtime
and the distribution of the monthly
activity degree. Our aim is to point out the main macroscopic
features of urban mobility studying their correlation with the
idea of "asystematic mobility" recently proposed by sociologists to
explain the observational data in modern metropolis. We consider
GPS data recorded during March 2008 in the Florence urban area. We show that some simple
assumptions on single particles, like the existence of an "individual
mobility energy" and of an "individual mobility time" that define the daily agenda,
may explain the statistical laws emerging from
the GPS data. Moreover in the equilibrium state individuals
seem to minimize their interactions, behaving independently, so
that the Maximum Entropy Principle of statistical mechanics can be
applied\cite{landau1980}. These results are consistent with the idea
that the sprawling phenomenon of modern cities implies that
citizens move as stochastic particles\cite{batty2005}. Finally
our analysis enlightens some average cognitive properties
of individuals in urban mobility. The paper is organized as follows: in the first
section we shortly described the GPS data base for vehicle mobility; in the remaining
sections we discuss the three statistical laws on path lengths distribution,
on the activities downtime and on the activity degree that are inferred from the data.

\section{GPS data for vehicle mobility}

A sample of $\simeq 2\%$ of the private vehicles in Italy has a GPS system for insurance reason.
Any vehicle is associated to an ID number, so that it is possible to follow its mobility during a long time.
Each datum gives position, velocity, covered distance from the previous measure and quality of signal.
The data give a sampling of individual trajectories each $\simeq 2$ km, but a signal is also recorded
any time the engine is switched on or off. The data suffer from the GPS limited
precision, in particular when the GPS looses the satellite signal. These problems are especially relevant
at starting points of the trajectories or when vehicles are parked inside a building, and
short paths could be strongly affected by these pathologies. When the quality of signal is good the time precision of the recorded
data is practically perfect, whereas the space precision is of the order of $10\, m$, usually sufficient
to localize a vehicle on the road. Both the instantaneous velocity and the covered space, are given with an
adequate precision since they result from a calculation based on GPS data recorded each second, but not registered.
We have developed several methods to clean the data from spurious effects in order to avoid passible bias
in the sample.\par
\begin{figure}[htbp]
\centerline{\includegraphics[width=6 truecm , bb= 20 20 575 575]{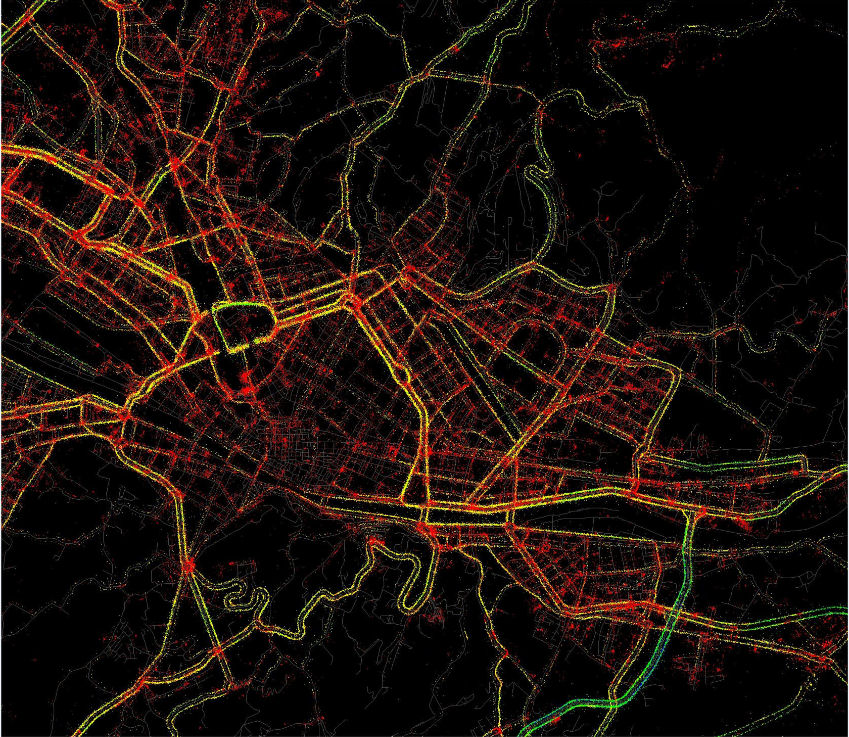}}
\caption{Aggregated GPS position data in the Florence area recorded
during March 2008; the red dots corresponds to a recorded
instantaneous velocity $\le 30$ Km/h, whereas the yellow dots
correspond to velocities in the interval $30\div 60$ Km/h and the
green dots to a velocity $\ge 60$ km/h.} \label{figure1}
\end{figure}

\noindent
In the present work we consider the GPS data in the Florence urban area recorded during March 2008: these data are related
to 35,000 vehicles in a circular area of radius $\simeq 30$ Km, around the historical center and defining $\simeq 2.5\times 10^6$
different trajectories. We have restricted our analysis to the trajectories which start inside a circle
of 10 Km around the historical town and remain inside the considered area, so that with a good probability we select
people living and moving in Florence.
Then we look for the vehicles which perform daily loops from starting points that we identify as "home": in this
case the number of trajectories reduces to $\simeq 4.\times 10^5$.  In figure \ref{figure1} we show
the considered area where we have plotted the aggregate position GPS data: the color refers to the different
instantaneous velocities (red means less than $30$ Km/h, whereas yellow refers to a velocity in the interval
$30\div 60$ Km/h and the green to a velocity $\ge 60$ km/h).
We are completely ignorant on the social composition of the sample and on the specific drivers, but we
expect that such individuals perform a mobility
related to the activities present in the Florence area and have a certain knowledge of the road network.

\section{Path length distribution}

\begin{figure}[htbp]
\centerline{\includegraphics[width=7 truecm , bb= 20 20 575 575]{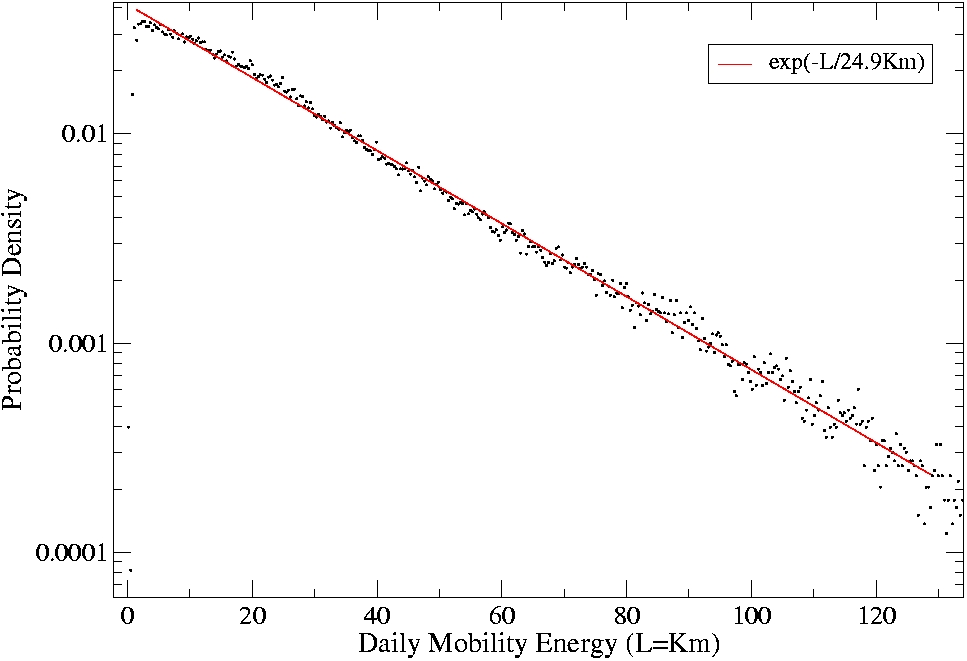}}
\caption{Empirical distribution of the daily mobility length $L$(dots).
The continuous line is an interpolation with a Maxwell-Boltzmann distribution with average value $<L>\simeq 25\, Km$.} \label{figure2}
\end{figure}

The activity sprawling that characterizes the modern metropolis has certainly a strong influence on the
individual mobility demand\cite{domencich1975}. Even if the Florence historical center is a very special area full of artistic and tourist attractions,
but forbidden to private traffic, nevertheless we assume the activities randomly distributed in the urban system.
This hypothesis is quite reasonable because we consider a large urban area and our sample
is surely composed by inhabitants and not tourists.
As a consequence, we expect that the citizen mobility agenda are
influenced by individual features rather than by the city structure. In cities
the stationary average traffic state should emerge as the result of individual interactions of cognitive
particles which share the same spatial resources. In particular we assume that drivers
organize their mobility, by applying a minimization strategy of the interactions with other individuals\cite{volkov2009}.
In this conceptual framework, the mobility can be seen as the
realization of many independent individual agenda and the dynamical properties become similar to that of a Boltzmann gas.
Even if it is obviously true that individuals are non-identical particles, path length and activity downtime can be
considered common mobility features to all people, and good candidates for a statistical physics approach to
describe the stationary state. For each vehicle we have recorded the total lengths $L$ of his daily round trips for the whole
considered period. The length distribution is plotted in fig. \ref{figure2} where we point out the existence of
an interpolation with a Maxwell-Boltzmann distribution
\begin{equation}
p(L)=L_0\exp(-L/L_0)
\label{(1.1)}
\end{equation}
with $L_0\simeq 25$ Km the characteristic daily path length.
The distribution (\ref{(1.1)}) provides a very good fit of the experimental data,
and it can be justified by the Maximum Entropy Principle under the assumption that the individuals are
independent particles and that there exists an average daily trip length in the population (see supplementary material). In such
a case the distribution (\ref{(1.1)}) is realized when any particle chooses its mobility energy randomly.
It is straightforward to associate an individual "mobility energy" to the daily path lengths.
The assumption that citizens organize their mobility as they own an internal "mobility energy",
agrees with similar hypotheses discussed by R.Kolb and D.Helbing\cite{kolb2003}
to explain the daily travel-time distributions for different transport modes.
\par
\begin{figure}[htbp]
\centerline{\includegraphics[width=7 truecm , bb= 20 20 575 575]{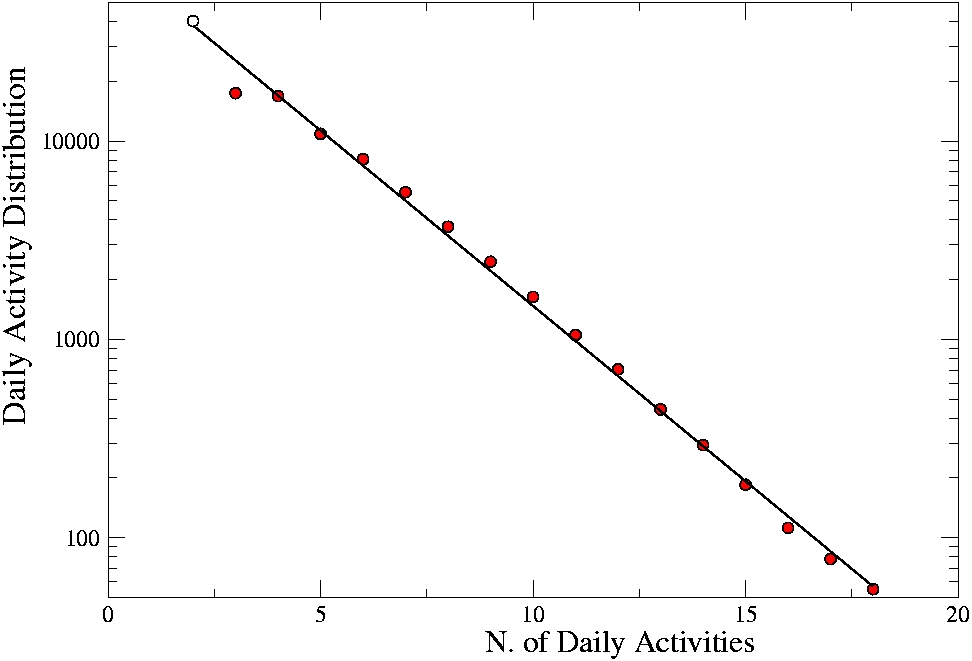}}
\caption{Rank distribution of daily activities computed from the GPS data(circles).
The continuous line is an exponential interpolation according to (2)
with $a\simeq .7$.} \label{figure3}
\end{figure}

\begin{figure}[htbp]
\centerline{\includegraphics[width=7 truecm , bb= 20 20 575 575]{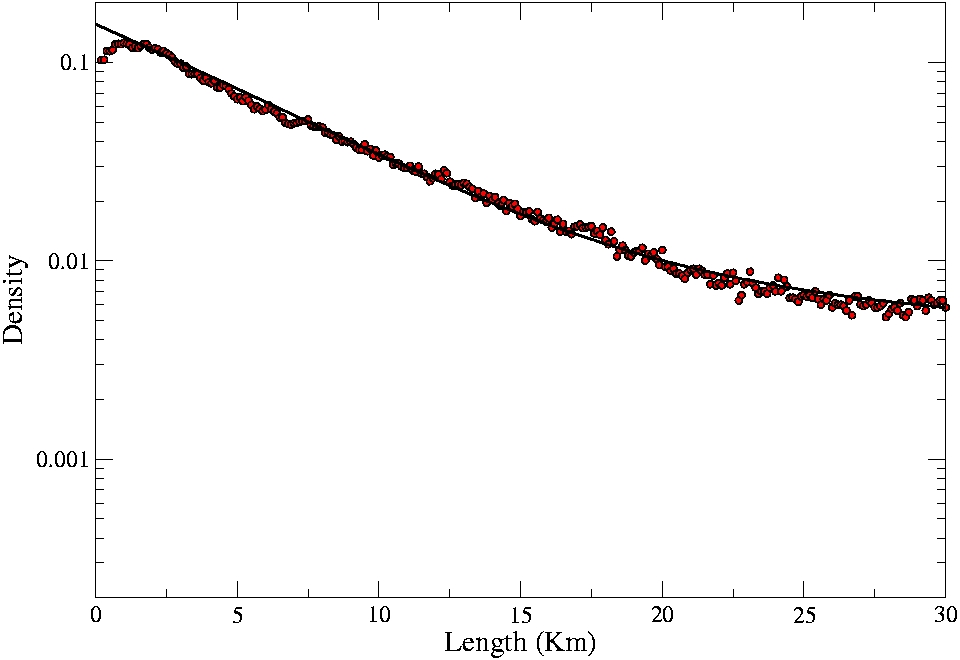}}
\caption{Single trip length distribution computed from the GPS data(circles).
The continuous line corresponds to an interpolation with the distribution (\ref{(2.2)})
by fixing the maximal number of daily activities at $N=18$ according to fig. \ref{figure3} and
an effective daily mobility length $L=30$ Km (not optimized).} \label{figure4}
\end{figure}

\noindent
In order to investigate the concept of the mobility energy together with the Maximum Entropy Principle, we consider the
relation between the daily path length and the single trip path length, building up the rank distribution of individual daily activities.
To define an activity from GPS mobility data, we apply a clustering procedure to
vehicle stop positions, identifying the positions which lie in a circle of diameter $\simeq 500$ m (this is considered a
acceptable distance between the true destination and the parking place\cite{benenson2008}). Moreover we have associated an activity when the
elapsed time before the next trip is greater than 15 minutes.
In figure \ref{figure3} we plot the rank distribution for the daily activities together with an exponential interpolation
\begin{equation}
p(k)\propto a^k
\label{(2.1)}
\end{equation}
that provides a very good fit of experimental observations. The equation (\ref{(2.1)}) is consistent with the
assumption that, in average, the individuals behave as independent random particles which
define their daily agenda in a random way.
According to previous hypotheses,  it is possible to compute in analytical way
the single trip length distribution, as the distribution realized by uniformly spreading $k$
points into a given segment of length $L$. A simple calculation provides the single trip length distribution in the form
(see appendix)
\begin{equation}
p_N(x)={c\over L}\sum_{k=1}^N(k+1)ka^k(1-x/L)^{k-1}
\label{(2.2)}
\end{equation}
where $c$ is a normalizing factor and $N$ is the maximum number of daily activities;
we remark that the choice of the points in the segment is contextual without any time-ordering. It is quite
natural to assume that there should exist a correlation between the number of daily activities $N$ and
the daily mobility length $L$, but the GPS data do not suggest any correlation function, so that we decided
to use an effective daily length in the theoretical distribution (\ref{(2.2)}) to
make a comparison with the empirical one. It turns out that If we exclude the very short paths, the curve (\ref{(2.2)}) fits very well with
the experimental data (see fig. \ref{figure4}) with an effective daily mobility length $L=30$ Km (the result is not
too sensitive to this particular value and we have not performed any optimization procedure).
As a final remark from the figure \ref{figure4}, we observe that the long trip length distribution
differs from an exponential behavior and it rather seems to follow a power law.

\section{Activity downtime distribution}

\begin{figure}[htbp]
\centerline{\includegraphics[width=7 truecm , bb= 20 20 575 575]{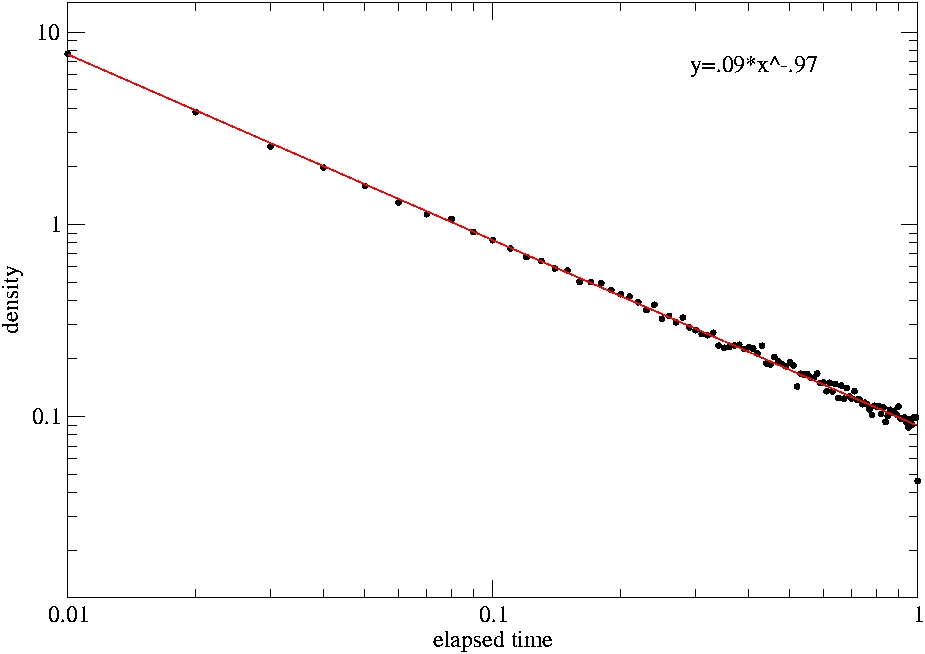}}
\caption{Activity downtime distribution computed from the GPS data(circles); we have normalized the data at the maximal value.
The continuous line corresponds to an interpolation with a power law $x^\alpha$ with $\alpha=.97$
which is consistent with the Benford's law (\ref{(3.1)}).} \label{figure5}
\end{figure}

Time is the second fundamental individual variable of human mobility, directly related
to the dynamical realization of daily agenda. From the GPS data base in the Florence area we have
computed the downtime spent in each daily activity by a fixed individual; we discard from the activities
the sleeping time linked to circadian rhythms.
The distribution of the activity downtimes of the recorded individuals is plotted
in fig. \ref{figure5}, recovering the well known Benford's law\cite{pietronero2001}.
In order to give a microscopical interpretation of the empirical downtime distribution, we assume
that individuals cannot determine a priori each activity downtime, because this is varied depending on
unpredictable circumstances. According to this hypothesis, in the average, each particle has a finite mobility time at disposal
to perform the desired activities, and he consumes the time in successive random choices up to the end of the whole mobility
time. If one computes the interval distribution that is obtained by choosing
successively $k$ points in a given segment, one get analytically the Benford's distribution\cite{pietronero2001}
\begin{equation}
p(t)\propto -\sum_{k=1}^N {(\ln t)^(k-1)\over (k-1)!}\simeq {1\over t}
\label{(3.1)}
\end{equation}
Shortly, the statistical results of the monthly mobility in the Florence area, recorded by the GPS data on vehicles,
suggest that the macroscopic average properties are the same of those of Boltzmann's particles moving in
a homogeneous space with an average energy and a finite time at disposal. The
energy introduces a global constraint in the individual daily mobility, whereas the time can be seen as a
local constraint in the activity planning since it is consumed step by step.\par\noindent
However computing the distribution of the total activity downtime for the monitored vehicles, we can
get an idea of the typology of urban mobility described by our sample.
\begin{figure}[htbp]
\centerline{\includegraphics[width=7 truecm , bb= 20 20 575 575]{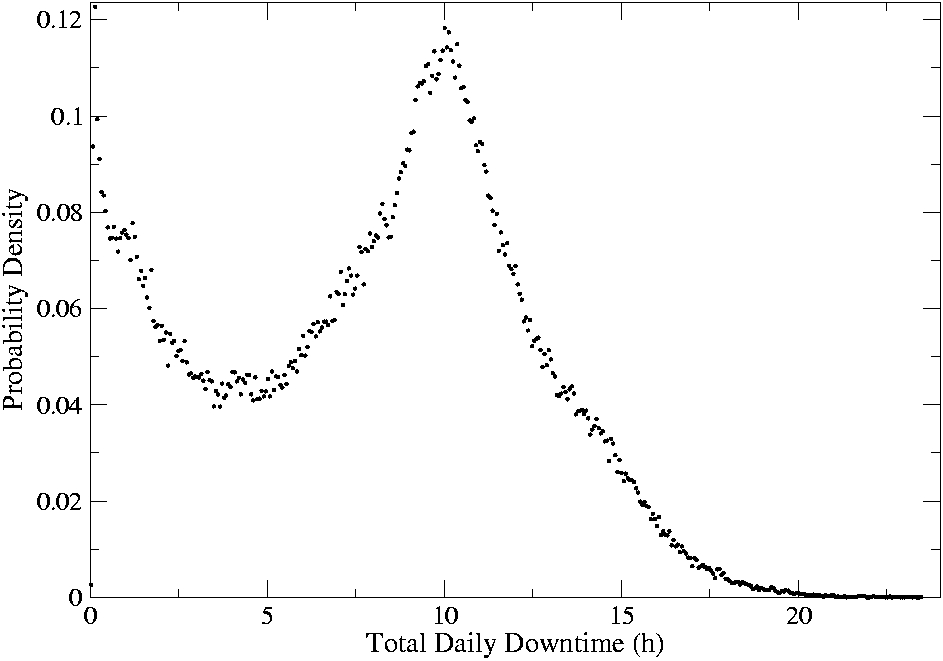}}
\caption{Daily activity downtime distribution.
We remark three characters of the underlying mobility: the first peak is related to short downtime activities ($\simeq 1$ h), the
second peak ($\simeq$ 10 h) is probably due to working people and the long tail ($\ge 12$ h) could be explained by the presence
of business vehicles that are used by different people.} \label{figure6}
\end{figure}
The results are plotted in fig. \ref{figure6}
We remark the presence of two peaks: one is related to short downtime activities, that probably corresponds to a specific use
of the vehicle for a single trip, whereas the second peak centered at $\simeq 10$ h denotes people performing a more complex mobility
agenda, which contains the working activities. Finally the presence of a long queue for the daily activity downtime ($\ge 12$ h) is
probably due to business vehicles in our sample, that are used by different people.

\section{Activity degree distribution}

Both the Boltzmann distribution for the mobility energy and the Benford's law for the activity
downtime enroll stochastic features of the system, but they do not explain how such features can be related
to the individual daily agenda, that are certainly the result of a cognitive behavior.
In order to study this question, we perform a statistical analysis of the downtime related to daily activities, considering
the monthly degree $k$ for the different individual activities (i.e. the number of times that a citizen repeats a certain activity
during a month)\cite{schonf2004}. Let $t$ the activity downtime, we introduce the join probability $p(t,k)$ to
denote the probability of finding a $k$-degree activity associated to a downtime $t$. Then by definition
we have to recover the Benford's law (\ref{(3.1)}) by summing over $k$
\begin{equation}
\sum_k p(t,k)\propto {1\over t}
\label{(5.1)}
\end{equation}
We have also the equality
$$
p(t,k)=p(t\mid k)k p(k)
$$
where $p(t\mid k)$ is the conditional probability for a downtime $t$ considering only the $k$-degree activities,
and $p(k)$ is the probability to detect a $k$ degree activity; the factor $k$ takes into account the multiplicity
of the $k$ degree activities.
The study of the conditional probability $p(t\mid k)$ can shed some light to understand the mobility habits related to the
use of private vehicles and to face the question of the relevance of repeated activities both in the mobility and in the
use of time. Remarkably the experimental observation suggest the existence of an universal probability distribution
$f(u)$ for the normalized downtime $t/<t>_k$:
\begin{equation}
p(t\mid k)={f(t/<t>_k)\over <t>_k}
\label{cond}
\end{equation}
where $<t>_k$ is the average downtime for the $k$-degree activities. We read this universal function as the signature of the fact that
individuals organize their time, when performing a private car mobility, in a common way independently
from the specific activity, i.e. the relative downtime fluctuations are the result of a stochastic universal mechanism.
Moreover there should exist a common feature among the individuals, concerning how they manage the downtime related to
the $k$-degree activities, since only the average value $<t>_k$ characterizes the $k$ dependence of the conditional
probability $p(t\mid k)$. This universal character could be explained thinking that the $t/<t>_k$ variable is
a generic "measure" of the mobility actions, valid for every individual. More precisely $t/<t>_k$ can be considered
the temporal norm for all the mobility related activities.
From the empirical data we detect $\simeq 3\times 10^5$ activity downtimes and we have computed the dependence of the average value $<t>_k$
using the degree $k=3,..,20$.
\begin{figure}[htbp]
\centerline{\includegraphics[width=7 truecm , bb= 20 20 575 575]{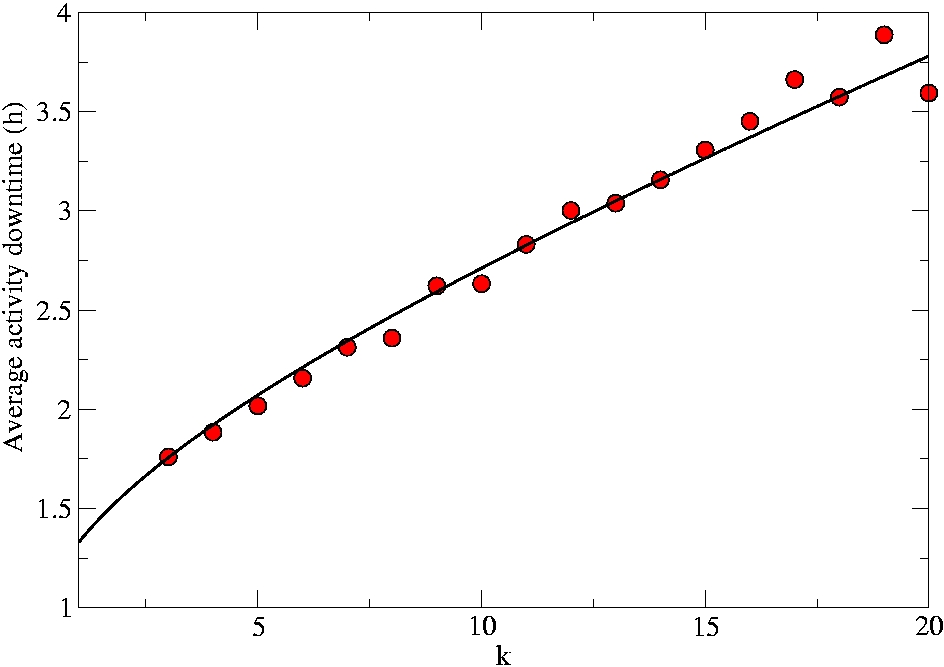}}
\caption{Dependence of the average downtime $<t>_k$ from the activities degree $k$.
The continuous line refer to a possible interpolation with the exponential function (\ref{inter}).} \label{figure7}
\end{figure}
It is evident from fig. \ref{figure7} that we have an almost linearly increasing behavior of $<t>_k$ as the degree $k$ increases. This means the existence of a relation
between the activity degree and the activity "use value" (individual satisfaction, profit, etc...) that introduces an individual tendency
to repeat and to spend time in the activities with a relevant added value\cite{samuleson2004}.
A possible local interpolation of the empirical data is obtained by using the
function (continuous line in fig. \ref{figure7})
\begin{equation}
<t>_k\propto \exp(\gamma k^a)
\label{inter}
\end{equation}
where $a\simeq .3$ and $\gamma\simeq .7$. In figures \ref{figure8},\ref{figure9} and \ref{figure10} we plot the empirical probability
densities for different degree (from $k=3$ to $k=20$) to investigate the existence of the universal distribution $f(u)$.
\begin{figure}[htbp]
\centerline{\includegraphics[width=7 truecm , bb= 20 20 575 575]{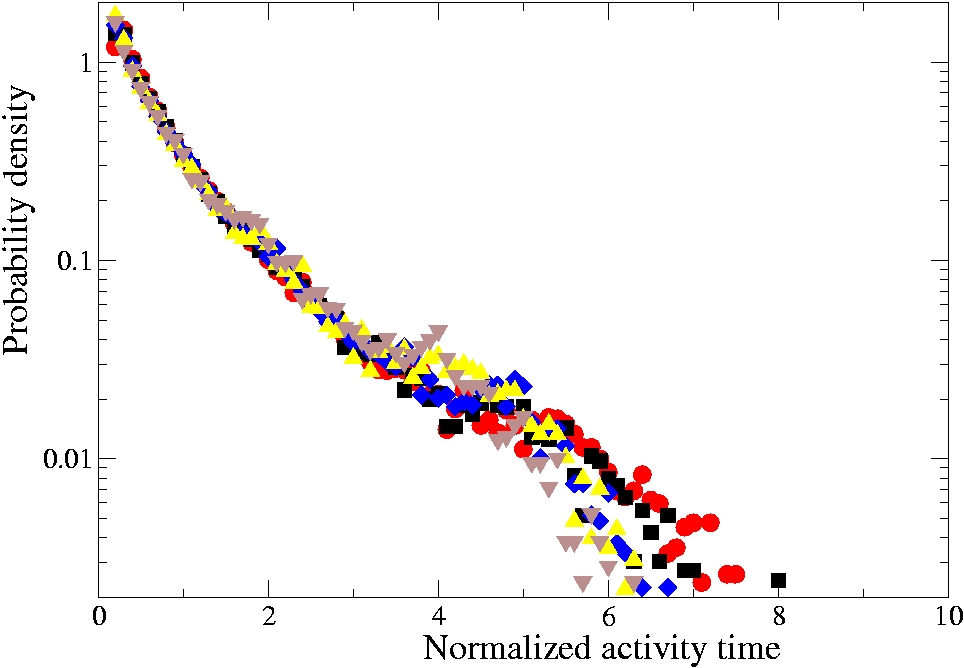}}
\caption{Empirical distributions of the $k$-degree activity downtime as a function of the normalized
downtime $t/<t>_k$. The different symbols
refer to the different activity degrees: $k=3$ (circles), $k=4$ (squares), $k=5$ (rhumbs), $k=6$ (up triangles),
$k=7$ (down triangles).} \label{figure8}
\end{figure}
\begin{figure}[htbp]
\centerline{\includegraphics[width=7 truecm , bb= 20 20 575 575]{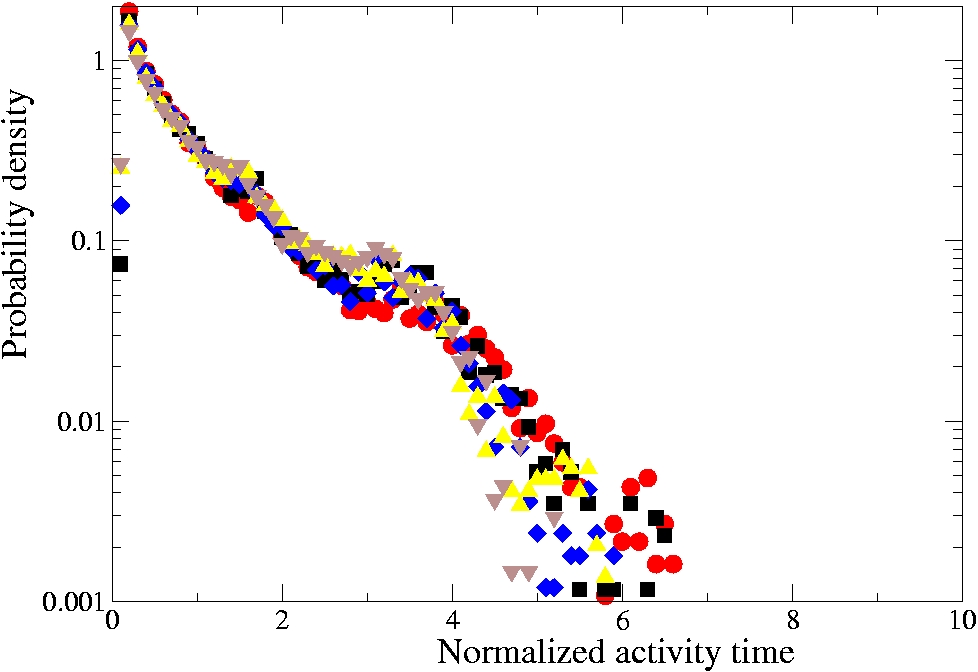}}
\caption{Empirical distributions of the $k$-degree activity downtime as a function of the normalized
downtime $t/<t>_k$. The different symbols
refer to the different activity degrees: $k=8$ (circles), $k=9$ (squares), $k=10$ (rhumbs), $k=11$ (up triangles),
$k=12$ (down triangles)..} \label{figure9}
\end{figure}
There is a decreasing of the data number as $k$ increases, but all the distributions
are computed with a sample of the same order (from $4\times 10^4$ to $10^4$).
The figures enlighten three different features. There is a collapse of all the curves on a unique distribution: this
is clear in the figure \ref{figure8} (the tail spread is consistent with statistical fluctuations) and in the first
part of all the plotted distributions that contains the great majority of the data. All the distributions show
a big contribution from the short times activities and a fast decaying tail for large $(t/<t_k>)> 2$.
\begin{figure}[htbp]
\centerline{\includegraphics[width=7 truecm , bb= 20 20 575 575]{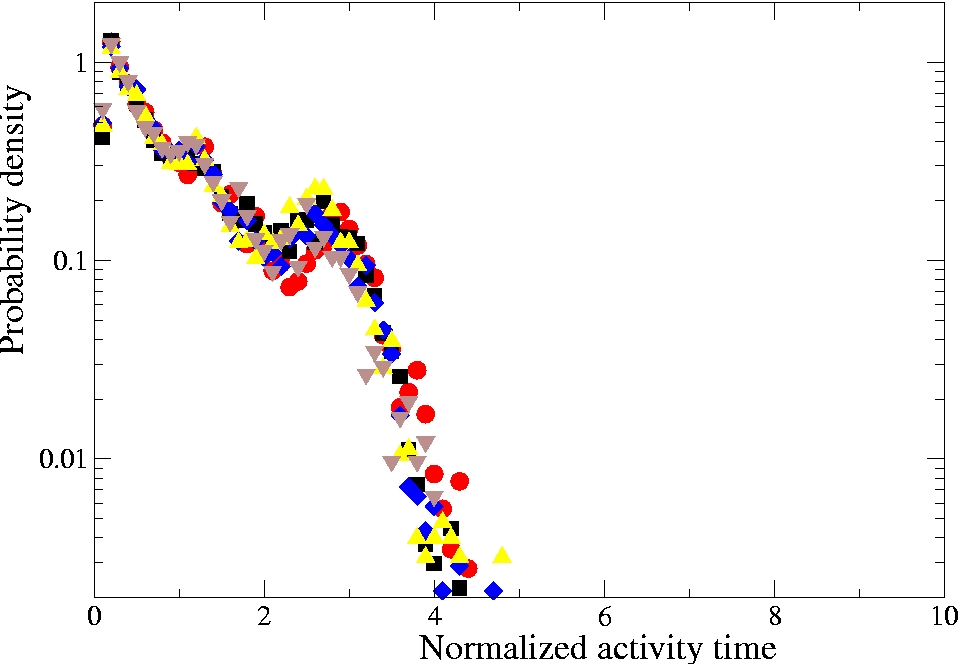}}
\caption{Empirical distributions of the $k$-degree activity downtime as a function of the normalized
downtime $t/<t>_k$. The different symbols
refer to the different activity degrees: $k=16$ (circles), $k=17$ (squares), $k=18$ (rhumbs), $k=19$ (up triangles),
$k=20$ (down triangles).} \label{figure10}
\end{figure}
There is a smooth rise of a "signal"
as $k$ increases denoted by the appearance of two peaks at $t/<t>_k\simeq 1,3$: this is clear in the last figure \ref{figure10}.
Therefore the empirical observation gives a strong indication for the existence of an universal distribution $f(u)$ for
the normalized activity downtime, even if when we consider high degree activities ($k\ge 10$) some new features appear but
with a small statistical weight. A possible interpolation of the distribution $f(u)$ is given by
\begin{equation}
f(u)\propto {1\over u}e^{-\alpha u}
\label{univ}
\end{equation}
where the coefficient $\alpha$ has a value $\simeq .4$. The distribution (\ref{univ}) is singular at the origin so that the
interpolation is certainly approximated at $u\to 0$ (see fig. \ref{figure3a} in the appendix).
\begin{figure}[htbp]
\centerline{\includegraphics[width=7 truecm , bb= 20 20 575 575]{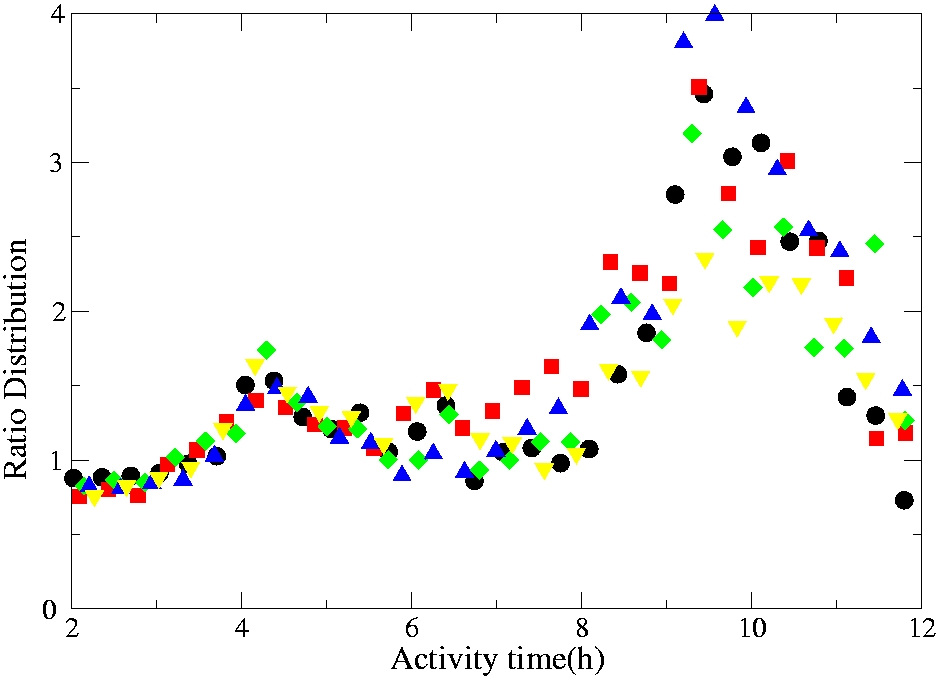}}
\caption{Ratio between the empirical distributions of the $k$-degree activity downtime and the interpolation
(\ref{inter}) as a function of time (h). The different symbols refer to the different activity degrees: $k=16$ (circles),
$k=17$ (squares), $k=18$ (rhumbs), $k=19$ (up triangles), $k=20$ (down triangles). Two peaks at a downtime $t\simeq 4$ h
and at $t\simeq 10$ h that are common to all the distributions can be observed.} \label{figure11}
\end{figure}
The exponential decay is the typical Boltzmann
statistics as for the path length distribution, whereas the $u^{-1}$ behavior is consistent with the Benford's law for the
downtime distribution. So that this "universal distribution" seems to mix both the features shown in fig. \ref{figure4} and \ref{figure5}.
To explain the singular $f(u)$ trend at the origin, we guess that it can be a sign of a strong free-will in the individual behavior
in the short time activity range. Surely for a more precise justification further study are required.
We can use the interpolation (\ref{univ}) to extract the signal from the
high degree activity distribution, by computing the ratio between the empirical distribution data and the interpolation (\ref{inter}); in the figure
\ref{figure11} we plot the ratio distribution results. As it can be seen, the empirical data define two peaks centered at $t\simeq 4$ hours and at
$t\simeq 10$ that are common to all the distributions when $k=16,...,20$. Even if these peaks are not statistically relevant,
they can be related to individuals that perform repeated activities linked to the canonical working time schedule.
Clearly the working time schedule introduces further constrains to the individual mobility agenda,
that are not taken into account by the universal distribution $f(u)$.
Now the existence of an universal distribution implies (cfr. eq. (\ref{cond}))
\begin{equation}
p(t,k)=f\left({t\over <t>_k}\right){k p(k)\over <t_k>}
\label{probt}
\end{equation}
Then using the interpolation (\ref{inter}) and performing the change of variable $u(k)=t/\exp(\gamma k^a)$,
we obtain that the Benford's
law (\ref{(5.1)}) implies a power law distribution for the activity degrees(see appendix)
\begin{equation}
p(k)\simeq {1\over k^{2-a}}
\label{degree}
\end{equation}
According to the estimate
(\ref{inter}), we expect  an exponent $\simeq -1.7$. In fig. \ref{figure12} we plot the empirical activity degree distribution
 with a numerical interpolation by a power
law $k^{-b}$; the data provide $b\simeq 1.6$ which is consistent with the analytical estimate (\ref{degree}).
\begin{figure}[htbp]
\centerline{\includegraphics[width=7 truecm , bb= 20 20 575 575]{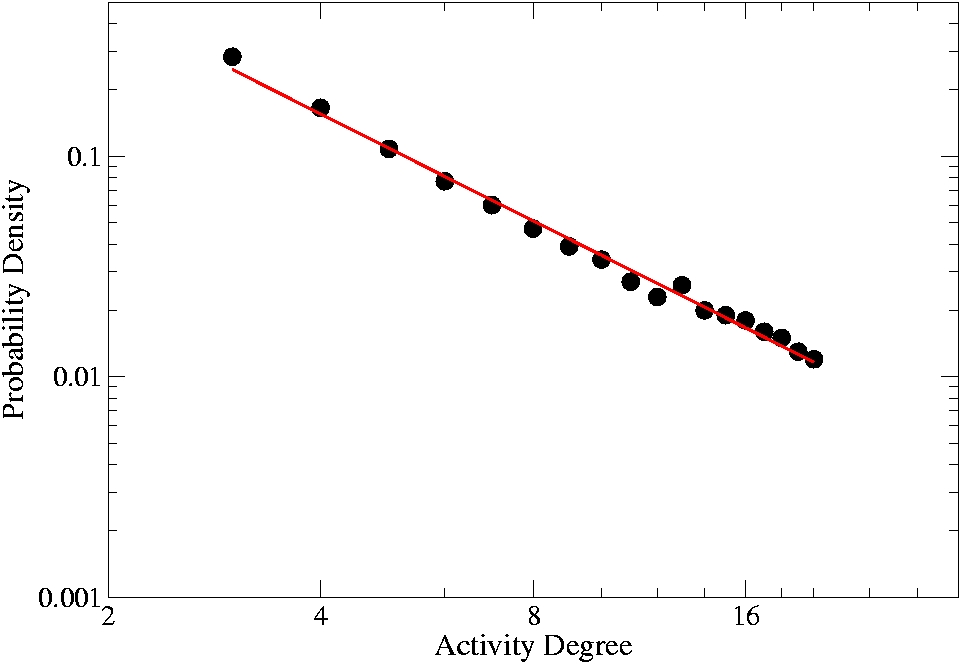}}
\caption{Empirical distributions of the activity degree (circles). In a log-log scale we enlighten the interpolation
with a power law $k^{-b}$ with $b\simeq 1.6$ (continuous curve).} \label{figure12}
\end{figure}

\section{Conclusions}

The citizens mobility is an interesting social phenomenon that involves a large number of
"intelligent" elementary components with a free
will individual property, so that we can speak of a cognitive dynamics. In this paper we
analyze a lot of car mobility data for the Florence area, showing the emergence of three
robust statistical laws for the path lengths, the activity downtime and degree.
These laws can be explained as the direct consequence of some simple
hypotheses as the particle independence, and above all assuming the existence of an
individual mobility energy and of a finite individual time spent for the desired urban
activities. Moreover, from the downtime distribution we deduce an universal
function which fits with empirical observations and data. We think that this universal
function can be interpreted as an indication of the cognitive time perception common to
all human beings, or at least surely common to the car drivers. Finally the
urban mobility is clearly complex, but the our steady state statistics cannot point out
the typical complexity signatures as, for instance, self-organized states. To detect
complexity in mobility dynamics, it is necessary to investigate the transients, i.e. the states far
from equilibrium.

\section{AKNOWLEDGEMENTS}
We thank OCTO Telematics for the access to the GPS data base on the Florence area. The authors are in debt with
Prof. Dirk Helbing and Prof. Luciano Pietronero for several stimulating discussions.

\section{Appendix}

\begin{figure}[htbp]
\centerline{\includegraphics[width=7 truecm , bb= 20 20 575 575]{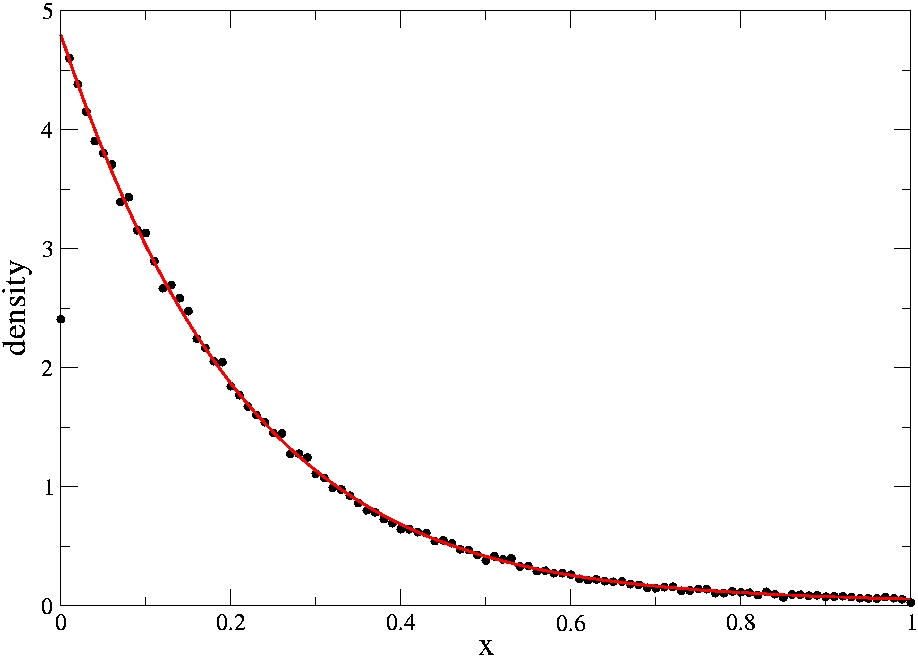}}
\caption{MonteCarlo simulations for the probability density of intervals in a given segment
defined by randomly choosing $n$ points in the interval uniformly distributed with $n$ a random integer
$\le N=7$. The continuous line is the plot of the theoretical distribution (\ref{a1}).} \label{figure2a}
\end{figure}

Let us consider $k$ stochastic variables uniformly distributed in the unit segment, the probability
that a segment of length $\le x$ is empty can be estimated according
$$
{\cal P}(\le x)=1-(1-x)^k
$$
As a consequence the probability density that a certain segment $x$ is empty is given by
$$
p(x)={d {\cal P}\over d x}=k(1-x)^{k-1}
$$
Therefore if one choices randomly an integer number $k$ in the interval $[1,N]$, the probability density
for a segment of length $x$ conditioned by the choice $k$ is
\begin{equation}
p_k(x)\propto(k+1)k(1-x)^{k-1}\qquad x\in[0,1]
\label{pk}
\end{equation}
since we have to take into account $k+1$ possible segments. The probability (\ref{pk}) has to be weighted by
the probability $p(k)\propto a^k$ to have $k$ points so that the probability density to detect
a segment of length $x$ for any choice $k$ is
\begin{equation}
p_N(x)={(1-a)^2\over (2-a)a(1-a^N)-Na^{N+1}(1-a)}\sum_{k=1}^N(k+1)k a^k(1-x)^{n-1}
\label{a1}
\end{equation}
where we have introduced a normalizing factor.
In fig. \ref{figure2a} we show the comparison between the equation (\ref{a1})
and a MonteCarlo distribution with $N=7$.
The distribution (\ref{a1}) allows an analytical approach to the single trip length distribution (cfr. fig. \ref{figure4} in
the paper).\par
\begin{figure}[htbp]
\centerline{\includegraphics[width=7 truecm , bb= 20 20 575 575]{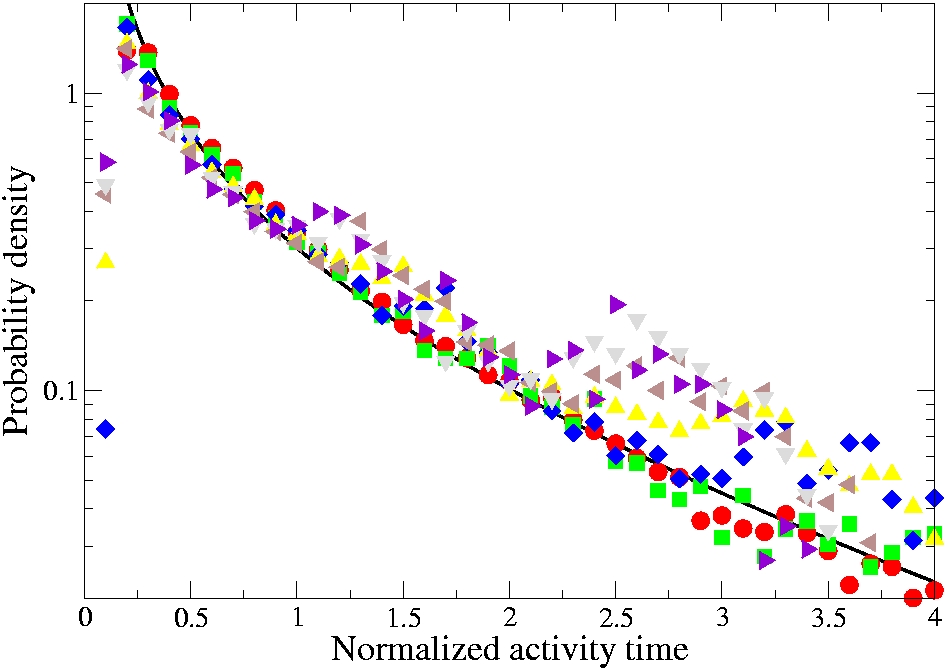}}
\caption{Empirical distributions for the conditional probabilities $p(t\mid k)$ for activities degrees $k=3,..,20$
as a function of the normalized activity downtime $t/<t>_k$ (different symbols). The continuous line refers to an interpolation
with the function (4).} \label{figure3a}
\end{figure}

\noindent
The empirical conditional distributions $p(t\mid k)$ for different degrees $k$
as a function of the normalized activity downtime $t/<t>_k$ suggests that the statistically relevant
can be described according to
\begin{equation}
p(t\mid k)={f(t/<t>_k)\over <t>_k}
\label {a5}
\end{equation}
where we introduce an universal function $f(t/<t>_k)$ which can be interpolated by
\begin{equation}
f(u)\propto {1\over u}e^{-\alpha u}
\label {a4}
\end{equation}
The results are shown in figure \ref{figure3a} for the activity degrees $k=3,..,20$.\par
There is a strict relation between the activity degree distribution (see fig. \ref{figure12} in the paper)
and the existence of an universal distribution probability $f(u)$ in eq. (\ref{a5}) Indeed taking advantage from
the dependence of $<t>_k$ on the degree $k$ pointed out by experimental observations (see fig. \ref{figure7} in the paper)
we perform the change of variables
\begin{equation}
\begin{cases}
t&=t\\
u&=t/<t>_k\\
\end{cases}
\end{equation}
in the join probability distribution $p(k,t)$ of degree and downtime (cfr. eq. (9)). Using the definition (5), we get the new distribution
$$
p'(u,t)=f(u)k(u)p(k(u)){d k\over du}=-f(u)kp(k){<t>_k\over t}\left ({d<t>_k\over dk}\right )^{-1}
$$
where $k$ has to be read $k(u)$ in the r.h.s. and $p(k)$ is the activity degree distribution. In the previous formula
we approximate interpolate the discrete variable $k$ with a continuous variable. By integrating of $u$ we have to recover
the Benford's law $\propto 1/t$ for the global activity downtime distribution (see fig. \ref{figure5} in the paper).
Since $f(u)$ is normalized as probability distribution, this is possible if
\begin{equation}
kp(k)<t>_k\left ({d<t>_k\over dk}\right )^{-1}={\rm const.}
\label {a6}
\end{equation}
According to the interpolation $<t>_k\propto \exp\gamma k^a$ of the experimental data as shown in the figure \ref{figure7}
in the paper, we explicitly have
$$
{d<t>_k\over dk}\propto k^{a-1}e^{\gamma k^a}\propto k^{a-1}<t>_k
$$
therefore the condition (\ref{a6}) reads
$$
k^{2-a}p(k)={\rm const.}
$$
i.e. a power law distribution of the activity degree with exponent $\le 2$. This is consistent with the experimental observations
as shown by the figure \ref{figure12} in the paper.

\begin{thebibliography}{10}


\bibitem{balescu1975}
R.~Balescu {\em Equilibrium and nonequilibrium statistical mechanics},
New York, Wiley-Interscience, (1975).

\bibitem{brockmann2005}
D.~Brockmann, L.~Hufnagel3 and T.~Geisel
{\em The scaling laws of human travel} Nature, \textbf{439}, (2006), pp. 462-465.

\bibitem{gonzalez2008}
M.C.~Gonz\'alez1, C.A.~Hidalgo and A.L.~Barab\'asi
{\em Understanding individual human mobility patterns} Nature, \textbf{453}, (5 June 2008), pp 779-782.

\bibitem{gelder1998}
T.van Gelder, {\em The dynamical hypothesis in cognitive science}, Behavioral and Brain Sciences \textbf{21}, (1998), pp. 615-628.

\bibitem{octotelematcs} http://traffico.octotelematics.it/.

\bibitem{bazzani2007}
A.~Bazzani, S.~Rambaldi, B.~Giorgini and L.~Giovannini {\em Mobility in modern cities: looking for physical laws}
ECCS07 Conference Proceedings, n. 132, (2007).
\bibitem{landau1980}
L.D.~Landau,E.M.~Lifshitz {\em Statistical Physics - Course of Theoretical Physics} \textbf{5}, Third edition, Butterworth-Heinemann, (1980)

\bibitem{batty2005}
M.~ Batty {\em Cities and Complexity} The MIT Press, Cambridge, Massachusetts
(2005).

\bibitem{domencich1975}
T.~Domencich and D.L.~McFadden
{\em Urban Travel Demand: A Behavioral Analysis}
North-Holland Publishing Co., (1975).

\bibitem{volkov2009}
I.~Volkov, J.R.~Banavar, S.P.~Hubbell and A.~Maritan {\em Infering species interactions in tropical forests} PNAS, \textbf{106-33},
(2009), pp. 13854-13859.

\bibitem{kolb2003}
R.~K\"olb and D.~Helbing {\em Energy laws in human travel behaviour}, New J. Phys. \textbf{5}, (2003) pp.48.1-48.12.

\bibitem{benenson2008}
I.~Benenson, K.~Martens {\em From modeling parking search to establishing urban policy}
Kunstliche Intelligenz, \textbf{3}, (2008), pp.3-8.

\bibitem{pietronero2001}
L.~Pietronero, E.~Tosatti, V.~ Tosatti and A.~Vespignani,  {\em
Explaining the uneven distribution of numbers in nature: The Laws of Benford and Zipf} Physica
A: Statistical Mechanics and its Applications, \textbf{293} (12), (2001), pp. 297-304.

\bibitem{schonf2004}
S.~Sch\"onfelder and K.W.~Axhausen {\em Structure and innovation of human activity space} Arbeitsbericht Verkerhrs und Raumplanung 258, IVT
and ETH Zurich, (2004).

\bibitem{samuleson2004}
P.~S.~Samuelson and W.~D.~Nordhaus {\em Economics 17th Edition}, McGraw-Hill, (2004).



\end{thebibliography}
\end{document}